\documentclass[sigconf]{acmart}

\usepackage{float}

\usepackage[utf8]{inputenc}

\usepackage{csquotes}

\usepackage{graphicx}
\graphicspath{{images/}}

\usepackage{hyperref}

\usepackage{xcolor}
\definecolor{deepblue}{rgb}{0,0,0.5}
\definecolor{deepred}{rgb}{0.6,0,0}
\definecolor{deepgreen}{rgb}{0,0.5,0}
\definecolor{commentgreen}{RGB}{2,112,10}
\definecolor{eminence}{RGB}{108,48,130}
\definecolor{weborange}{RGB}{255,165,0}
\definecolor{frenchplum}{RGB}{129,20,83}

\usepackage{listings}
\usepackage{pythonhighlight}
\usepackage{array}

\usepackage{multirow}

\usepackage{colortbl}

\usepackage{tcolorbox}

\copyrightyear{2022}
\acmYear{2022}
\acmConference[MSR 2022]{MSR '22: Proceedings of the 19th International Conference on Mining Software Repositories}{May 23–24, 2022}{Pittsburgh, PA, USA}

\begin{document}

% \title[Do TypeScript Applications Exhibit Better Software Quality than JavaScript Applications?]{Do TypeScript Applications Exhibit Better Software Quality than JavaScript Applications? A Repository Mining Study on GitHub}
\title[To Type or Not to Type? Comparing the Software Quality of JavaScript and TypeScript Applications]{To Type or Not to Type? A Systematic Comparison of the Software Quality of JavaScript and TypeScript Applications on GitHub}

\author{Justus Bogner}
\email{justus.bogner@iste.uni-stuttgart.de}
\orcid{0000-0001-5788-0991}
\affiliation{
  \institution{University of Stuttgart, Institute of Software Engineering}
  \city{Stuttgart}
  \country{Germany}
}

\author{Manuel Merkel}
\email{st155131@stud.uni-stuttgart.de}
\affiliation{
  \institution{University of Stuttgart, Institute of Software Engineering}
  \city{Stuttgart}
  \country{Germany}
}


\begin{abstract}
    JavaScript (JS) is one of the most popular programming languages, and widely used for web apps, mobile apps, desktop clients, and even backend development.
    Due to its dynamic and flexible nature, however, JS applications often have a reputation for poor software quality.
    While the type-safe superset TypeScript (TS) offers features to address these prejudices, there is currently insufficient empirical evidence to broadly support the claim that TS applications exhibit better software quality than JS applications.
    
    We therefore conducted a repository mining study based on 604 GitHub projects (299 for JS, 305 for TS) with over 16M LoC.
    Using SonarQube and the GitHub API, we collected and analyzed four facets of software quality: a) code quality (\# of code smells per LoC), b) code understandability (cognitive complexity per LoC), c) bug proneness (bug fix commit ratio), and d) bug resolution time (mean time a bug issue is open).
    For TS, we also collected how frequently the type-safety ignoring \texttt{any} type was used per project via ESLint.
    
    The analysis indicates that TS applications exhibit significantly better code quality and understandability than JS applications.
    Contrary to expectations, however, bug proneness and bug resolution time of our TS sample were \textit{not} significantly lower than for JS: the mean bug fix commit ratio of TS projects was more than 60\% larger (0.126 vs. 0.206), and TS projects needed on average more than an additional day to fix bugs (31.86 vs. 33.04 days).
    Furthermore, reducing the usage of the \texttt{any} type in TS apps appears to be beneficial: its frequency was significantly correlated with all metrics except bug proneness, even though the correlations were of small strengths (Spearman's rho between 0.17 and 0.26).
    
    Our results indicate that the perceived positive influence of TypeScript for avoiding bugs in comparison to JavaScript may be more complicated than assumed.
    While using TS seems to have benefits, it does not automatically lead to less and easier to fix bugs.
    However, more research is needed in this area, especially concerning the potential influence of project complexity and developer experience.
\end{abstract}




% For page numbers at the bottom
\thispagestyle{plain}
\pagestyle{plain}

\maketitle

\section{Introduction}
In 1995, JavaScript (JS) was introduced to add dynamic client-side functionality in the web browser~\cite{Flanagan2020}.
With the growth of the Internet, its usage spread rapidly.
According to a recent Stackoverflow survey\footnote{\url{https://insights.stackoverflow.com/survey/2021\#section-most-popular-technologies-programming-scripting-and-markup-languages}}, JavaScript is now the most widely used programming language.
Potential reasons may include its versatility and ease of use.
TypeScript, a superset of JavaScript, is also growing in popularity\footnote{\url{https://octoverse.github.com/\#top-languages-over-the-years}}, with some claiming it will be \textit{the} programming language for developing next-generation web apps, mobile apps, Node.js apps, and IoT software~\cite{Cherny2019}.

TypeScript (TS) mainly extends JavaScript with type annotations and provides \texttt{tsc}, the TypeScript Compiler to transpile TS to JS.
TS performs type checking at transpile-time, while JS performs type checking at run-time~\cite{Jansen2015}.
This is one of the reasons why JavaScript does not have the best reputation for delivering high-quality software.
Its beginner-friendliness and dynamic nature without a compiler lead some people to assume that using JavaScript often leads to poor software quality~\cite{Fard2013,Roehm2019}, similar to sentiments about other scripting languages like PHP~\cite{Amanatidis2016}.
However, there is currently insufficient empirical evidence to support the claim that TypeScript leads to overall better software quality than JavaScript.
The debate whether statically typed languages are better for software quality than dynamically typed ones has been going on for quite some time.
Proponents of static typing insist that \enquote{it allows early detection of some programming errors}~\cite{Pierce2002} and that it leads to \enquote{better documentation in the form of type signatures}~\cite{Meijer2004}.
Advocates of dynamic typing, however, claim that \enquote{it was easier to get things right with short source code, in which code that was not too terse or verbose determined behavior, when all types could be coerced into strings for debugging}~\cite{Loui2008} and that \enquote{static typing is too rigid, and that the softness of dynamically [\textit{sic}] languages makes them ideally suited for prototyping systems with changing or unknown requirements}~\cite{Meijer2004}.

Several studies have tried to empirically analyze the impact of static typing on software quality~\cite{Hanenberg2010,Kleinschmager2012,Ray2014,Gao2017,Zhang2020}, unfortunately with inconclusive or contradictory results (see Section~\ref{sec:relWork}).
% think about adding Prechelt1998
Moreover, many of the existing repository mining studies suffer from different threats to validity, making their comparison and replication difficult~\cite{Hassan2008,Robles2010}.
To the best of our knowledge, there is also no study that directly analyzes and compares JavaScript and TypeScript applications in terms of software quality on a large scale.
The aim of this study is therefore to verify the claim that TypeScript applications exhibit better software quality.
To do so, we collected and analyzed a large set of JavaScript and TypeScript applications on GitHub, a total of 604 projects with over 16M LoC, and compared different facets of software quality between them.

\section{Background}
In this section, we discuss the necessary foundations for this study, i.e., JavaScript, TypeScript, and their different type systems.

\subsection{JavaScript}
JavaScript (JS) is a scripting language developed by Netscape in 1995 for dynamic functionality in web browsers~\cite{Flanagan2020}.
JS enables rich interactions in web user interfaces and is especially popular for developing single page applications (SPAs)~\cite{Scott2015}.
The interpreted programming language is dynamically typed, and supports imperative, object-oriented, as well as functional programming.
However, this dynamic and flexible nature encouraged the prejudice that many JavaScript applications would show poor software quality~\cite{Fard2013,Saboury2017,Roehm2019}.
To harmonize JavaScript usage in different execution environments, Ecma International provides standardized ECMAScript (ES) versions\footnote{\url{https://www.ecma-international.org/technical-committees/tc39/}}.
A new version is released every year, with features like classes, modules, or array functions, thereby enhancing JavaScript further for building more complex applications.
Today, all modern browsers support the majority of ECMAScript features, especially for older versions like ES5 or ES6.
Through the introduction of Node.js~\cite{Tilkov2010} and Electron.js\footnote{\url{https://www.electronjs.org}}, JavaScript usage also spread beyond the browser, adding to its world-wide popularity.

\subsection{TypeScript}
TypeScript (TS) is a superset of JavaScript, and was released by Microsoft in 2012\footnote{\url{https://www.typescriptlang.org}}.
Superset means that it uses all ECMAScript features plus additional ones, especially an explicit type system with types such as \texttt{number}, \texttt{boolean}, \texttt{string}, \texttt{Array}, \texttt{Enum}, or custom \texttt{interface}s.
In general, valid ECMAScript is therefore also valid TypeScript.
TypeScript is statically typed and transpiled~\cite{Fain2020}, i.e., the source code is converted to JavaScript with the TypeScript compiler (\texttt{tsc}).
In addition to other analyses, \texttt{tsc} performs type checking during this conversion, and reports type-related bugs.
However, type annotations are optional: when we do not know the concrete types or deliberately want to keep them unspecified, the \texttt{any} type\footnote{\url{https://www.typescriptlang.org/docs/handbook/2/everyday-types.html\#any}} can be used.
A common use case for this is the migration of existing JS projects to TS, where \texttt{any} types are gradually replaced with concrete typings.
However, we lose type safety with the \texttt{any} type, and Microsoft therefore discourages its frequent permanent usage in an application.

\subsection{Type Systems}
A type system is a set of rules that assigns and enforces types to variables, functions, or objects in a programming language~\cite{Riscutia2019}.
The type defines what these entities represent, their set of allowed values, and which operations can be applied to them.
Developers either explicitly specify types or they are inferred from code context, such as variable assignments.
Type checking, i.e., ensuring that programs respect type system rules, is either done at compile time or during code execution.
\textit{Static typing} checks types at compile time, which then guarantees correct types at runtime.
\textit{Dynamic typing} only checks types during execution, which can lead to runtime errors due to incorrectly assigned types.
What this means for JavaScript as a dynamically typed language is illustrated in Fig.~\ref{lst:JavaScriptCode}.

\begin{figure}[ht]
\begin{lstlisting}
let myString = "I am a string";
myString = 42; // myString becomes a number

let myObject = {};
myObject.foo; // evaluates to undefined

function square(x) {
    return x * x;
}
square("a");  // evaluates to NaN (Not a Number)
square(true); // evaluates to 1
\end{lstlisting}
\vspace{-3mm}
\caption{JavaScript dynamic typing}
\label{lst:JavaScriptCode}
\vspace{-2mm}
\end{figure}

Instead of throwing an error, JavaScript handles not matching or missing types in these examples very liberally, and produces results that may be hard to understand and debug.
An equivalent example implemented with TypeScript is shown in Fig.~\ref{lst:TypeScriptCode}.

\begin{figure}[ht]
\begin{lstlisting}
let myString: string = "I am a string";
myString = 42;
// Type "number" is not assignable to type "string"

let myObject = {};
myObject.foo;
// Property "foo" does not exist on type "{}"

function square(x: number): number {
    return x * x;
}
square("a");
// Argument of type "string" is not assignable to
// parameter of type "number"
square(true);
// Argument of type "boolean" is not assignable to
// parameter of type "number"
\end{lstlisting}
\vspace{-3mm}
\caption{TypeScript static typing}
\label{lst:TypeScriptCode}
\vspace{-2mm}
\end{figure}

For all examples, \texttt{tsc} produces compilation errors with helpful error messages, which can also be displayed during development in editors or IDEs with TypeScript support.
Ultimately, both type systems have advantages and disadvantages, and it is important to decide which one is more suitable for the task at hand.
For smaller applications, the flexibility of dynamic typing could be preferable, while developing larger systems might be more sustainable with static typing.

\section{Related Work}
\label{sec:relWork}
% think about adding \cite{Prechelt1998}
% think about adding https://doi.org/10.1145/2936313.2816720
While we could only identify one study on the influence of TypeScript on software quality, several publications focus on quality differences between static and dynamic typing or between various programming languages.
Hanenberg~\cite{Hanenberg2010} conducted a controlled experiment with 49 undergraduate students, who had to develop a parser.
For the experiment, he created a new programming language, with similarities to Java, Ruby and Smalltalk, with a statically and a dynamically typed version. Participants were assigned to one of these two versions.
Hanenberg analyzed the required time for a minimal set of tests to pass and the resulting parser quality, which he evaluated with all 400 tests.
Contrary to his expectations, using the static type system did not have a significant positive impact on the programming time and quality.
For small tasks, there was no significant difference between the two groups, but when correcting type errors in the parser, the statically typed group even required significantly more time.

In a different experiment, Kleinschmager et al.~\cite{Kleinschmager2012} examined the development time for a programming task with a dynamically typed language, Groovy, and a statically typed language, Java.
A counter-balanced within-subject design was used, i.e., each of the 36 participants used both the static and the dynamic type system.
Three types of programming tasks were investigated: a) using existing code with documentation in the source code, b) fixing type errors in existing code, and c) fixing semantic errors in existing code.
As a result, Kleinschmager et al. found a positive influence of the static type system for six of the nine tasks, namely four of the five tasks of type a) and both tasks of type b).
For fixing semantic errors, however, no influence of the type system was found.
The result that fixing type errors was faster with the static type system is in conflict with the results of Hanenberg~\cite{Hanenberg2010}.

Ray et al.~\cite{Ray2014} performed a large-scale repository mining study to examine the effect of different programming languages on software quality.
They investigated the top programming languages from GitHub, which included JavaScript and TypeScript.
For each language, the top 50 repositories, sorted by stars, were examined, in total 850 projects in 17 languages.
They used a mixed-methods approach to analyze a) whether some languages are more defect prone than others and b) which language properties influence defect proneness.
Based on their analysis, they concluded \enquote{that static typing is better than dynamic}.

However, when Berger et al.~\cite{Berger2019} tried to replicate the results from Ray et al., they partially failed to do so: two of the four research questions could not be replicated due to missing code and irreconcilable differences in the data.
For example, Berger et al. had to remove TypeScript from the data set, as a large part of these projects originated from the period before 2012, i.e., predating the release of TypeScript.
Of the 41 repositories labeled as TS, only 16 really contained TS.
Furthermore, the classification of characteristics of the programming languages was mislabeled in several areas.
Berger et al. highlighted multiple threats to validity and stated that the conclusion of the original study would not hold.

In a more recent mining study, Zhang et al.~\cite{Zhang2020} examined a total of 600 GitHub projects, 60 for each of the 10 most popular programming languages.
The focus of their analysis were bug-fixing characteristics of different languages.
They compared bug resolution time between individual languages, as well as between static and dynamic typing.
Within their sample, dynamically typed languages consumed 59.5\% more time for bug resolution than statically typed ones.

Another large-scale mining study was conducted by Roehm et al.~\cite{Roehm2019}.
Using 6,897 GitHub projects for C, C++, C\#, Java, and JavaScript, they examined if common maintainability prejudices for these languages could be empirically verified.
Using the static analysis tool ConQuat, they collected five language-independent maintainability metrics: clone coverage, comment incompleteness, too long files, too long methods, and nesting depth.
As one of the results, they concluded that \enquote{JavaScript code is not worse than other code}.

One of the few studies solely focusing on TypeScript and software quality was conducted by Gao et al.~\cite{Gao2017}, with the goal to investigate whether real-world JavaScript bugs could have been prevented by static type checking.
They examined the bug fix commits and issue tracking data of 398 JavaScript projects from GitHub.
As a result, they found that using TypeScript would have prevented 60 of the 400 bugs they examined (15\%).

In summary, none of the existing studies compare the software quality of TypeScript and JavaScript applications at a large scale, and the results of studies on dynamic vs. static typing are inconclusive.
Gao et al. provide the first interesting empirical evidence, but there is no direct comparison to TS applications.
We therefore want to add more empirical evidence with a repository mining study that compares a substantial number of JS and TS applications.

\section{Research Design}
\label{sec:researchDesign}
With this study, we want to answer the following research questions:

\begin{enumerate}
    \item [\textbf{RQ{$_1$}}] Do TypeScript applications exhibit better software quality than JavaScript applications?
    \item [\textbf{RQ{$_2$}}] Do TypeScript applications that less frequently use the \texttt{any} type exhibit better software quality?
\end{enumerate}

For RQ1, we analyzed several facets of software quality to see if TypeScript apps perform significantly better.
For RQ2, we additionally analyzed if these facets of software quality are correlated with the number of used \texttt{any} types in TS projects, i.e., if there is a relation between the extent of type safety and software quality.
Both RQs are well suited for a mining software repository (MSR) study~\cite{Hassan2008,Kalliamvakou2015}, as open-source platforms like GitHub\footnote{\url{https://github.com}} provide large quantities of the necessary data for such analyses.
For transparency and reproducibility, we share our data and scripts using Zenodo\footnote{\url{https://doi.org/10.5281/zenodo.5886595}}.

\begin{table*}[ht]
    \centering
    \small
    \begin{tabular}{l p{2.7cm} p{6.7cm}  p{6.7cm}}
        RQ & Metric & Null Hypothesis & Alternative Hypothesis\\
        \hline
        \hline
        \multirow{8}{*}{RQ1}
        &
        Code smells per LoC
        &
        \textbf{H$_{0}^{1}$:} TS applications exhibit less or equal code quality than JS applications.
        & 
        \textbf{H$_{1}^{1}$:} TS applications exhibit better code quality than JS applications.\\
        & Cognitive complexity per LoC 
        & 
        \textbf{H$_{0}^{2}$:} TS applications exhibit less or equal code understandability than JS applications.
        &
        \textbf{H$_{1}^{2}$:} TS applications exhibit better code understandability than JS applications.\\
        & Bug fix commit ratio
        &
        \textbf{H$_{0}^{3}$:} TS applications are more or equally prone to bugs than JS applications.
        & 
        \textbf{H$_{1}^{3}$:} TS applications are less prone to bugs than JS applications.\\
        & Bug resolution time
        &
        \textbf{H$_{0}^{4}$:} TS applications require more or equal bug resolution time than JS applications.
        & 
        \textbf{H$_{1}^{4}$:} TS applications require less bug resolution time than JS applications.\\
        \hline
        \multirow{8}{*}{RQ2}
        &
        Code smells per LoC
        &
        \textbf{H$_{0}^{5}$:} The \texttt{any} type frequency correlates not or positively with code quality in TS applications.
        & 
        \textbf{H$_{1}^{5}$:} The \texttt{any} type frequency correlates negatively with code quality in TS applications.\\
        & Cognitive complexity per LoC     
        & 
        \textbf{H$_{0}^{6}$:} The \texttt{any} type frequency correlates not or positively with code understandability in TS applications.
        &
        \textbf{H$_{1}^{6}$:} The \texttt{any} type frequency correlates negatively with code understandability in TS applications.\\
        & Bug fix commit ratio
        &
        \textbf{H$_{0}^{7}$:} The \texttt{any} type frequency correlates not or negatively with bug proneness in TS applications.
        & 
        \textbf{H$_{1}^{7}$:} The \texttt{any} type frequency correlates positively with bug proneness in TS applications.\\
        & Bug resolution time
        &
        \textbf{H$_{0}^{8}$:} The \texttt{any} type frequency correlates not or negatively with bug resolution time in TS applications.
        & 
        \textbf{H$_{1}^{8}$:} The \texttt{any} type frequency correlates positively with bug resolution time in TS applications.\\
        \hline
    \end{tabular}
    \caption{Null hypotheses with their alternatives for both RQs}
    \label{tab:hypotheses}
\end{table*}

There are currently more than 300,000 JavaScript and 60,000 TypeScript repositories with five or more stars on GitHub.
In this study, we specifically wanted to analyze \textit{applications}, i.e., custom web, desktop, or smartphone end-user applications with JavaScript or TypeScript as the primary language.
Repositories containing, e.g., solely frameworks (like Angular\footnote{\url{https://github.com/angular/angular}}) or build tools (like Webpack\footnote{\url{https://github.com/webpack/webpack}}) were excluded.
For the selected projects, we focus on the following four frequently used characteristics of software quality:

\textbf{Code quality:}
Frequently used instruments to evaluate code quality are \textit{code smells}, i.e., indicators of low code quality which may impact maintainability~~\cite{Yamashita2013}.
Detecting and removing code smells can also prevent bugs~\cite{Emden2012}.
To obtain the number of code smells per project, we used the static analysis tool SonarQube\footnote{\url{https://www.sonarqube.org}}.
The number of code smell rules implemented in SonarQube is fairly balanced between the two languages, with 141 rules for JS and 147 for TS.
A large majority of the rules is also shared.
To eliminate project size as a confounder, we operationalized code quality for an individual project as the \textit{\# of code smells per LoC}.

\textbf{Code understandability:}
A more specialized facet of code quality is the degree to which developers can easily understand the code base.
A frequently used metric for this is \textit{cognitive complexity}, which was developed as a human-centric understandability metric to address short-comings of cyclomatic complexity~\cite{Campbell2018}.
The metric is already widely used in industry via SonarQube, and first empirical evaluations for its effectiveness look promising~\cite{Munoz2020}.
Similar to code smells, we operationalized code understandability of a project as \textit{cognitive complexity per LoC}, both collected via SonarQube.

\textbf{Bug proneness:}
The frequency of bugs is an often used proxy for functional correctness.
We queried the GitHub API for the entire commit history of a project to count the \textit{bug fix commits}.
These commits were identified by a script via commit messages containing \enquote{bug} and \enquote{fix}.
According to Zhang et al.~\cite{Zhang2020}, this is the most accurate method, with a precision around 95\%.
Manual examination of 500 commits classified as bug fixes confirmed this estimate in our sample.
Therefore, we operationalized bug proneness as the \textit{bug fix commit ratio}, i.e., for a given project, the number of bug fix commits is divided by the total number of commits.

\textbf{Bug resolution time:}
The time a bug issue is open is often used as an indicator for its complexity.
This duration can be easily determined via the GitHub issue tracking system\footnote{ \url{https://guides.github.com/features/issues/}}.
Since not every issue describes a bug, only those with a bug label or those containing \enquote{bug} in the title or description were included.
We added the latter because Bissyandé et al.~\cite{Bissyande2013} mentioned that only 30\% of all issues in their dataset had a label.
After manually checking our sample, however, it turned out that around 70\% of our bug issues have a label.
Following Zheng et al.~\cite{Zheng2015}, we operationalized bug resolution time as the \textit{mean duration from bug issue opening until the last issue comment} for a given project, as this would be more accurate than the closing of the issue.

Using these operationalized quality characteristics, we defined eight hypotheses for statistical analysis, four per RQ.
The null hypotheses H$_{0}^{1}$ to H$_{0}^{8}$ and their corresponding alternative hypotheses H$_{1}^{1}$ to H$_{1}^{8}$ are listed in Table~\ref{tab:hypotheses}.
The motivation behind the RQ1 hypotheses was that we assume TypeScript applications to have better overall software quality than JavaScript applications, since the latter have a reputation for bad software quality.
Reasons for this include that JS is easy to learn and thus popular for beginners, and that it is an interpreted and dynamically typed language without a compiler warning about erroneous or unoptimized code.
TypeScript, on the other hand, finds several types of errors at compile time, thereby presumably making it less bug prone.
Moreover, it is assumed that static typing enhances the self-documenting nature of code, which may make it easier to find and resolve bugs.
For RQ2, we assume that there is a correlation between using the \texttt{any} type less and improved software quality in TS projects, as frequent usage of the \texttt{any} type degrades type safety.

\subsection{Sampling}\label{sec:Sampling}
We used the official GitHub API to identify and collect relevant projects, as it  provides all necessary data.
A Python script was created to automate the sampling as much as possible.
However, a significant disadvantage of the GitHub API is that an authenticated user with an access token can only send 5,000 requests per hour.

Only repositories with more than five stars were selected, since projects below this usually have little or no activity~\cite{Borges2016}.
Furthermore, we also excluded forks.
In addition, only projects created between 2012 and mid-2021 were examined, as TypeScript was released in 2012.
Before 2012, the file extension \texttt{.ts} refers to XML files that contain human language translations~\cite{Berger2019}.
Lastly, the primary programming language (JS or TS) had to account for at least 60\% of the project.
Applications usually contain additionally languages such as HTML, CSS, or a backend language.
The code of the web framework Vue.js is listed on GitHub as a separate programming language.
Therefore, these projects were also examined.

Following Israel~\cite{Israel1992} to roughly estimate the required sample sizes for TS and JS, we combined a confidence level of 95\%, i.e., the probability that the selected data reflect the population, a level of precision of 5\%, i.e., the range at which the population data deviates from the samples, and an estimated population size.
As mentioned, there are more than 300,000 JavaScript and more than 60,000 TypeScript projects on GitHub which are potentially relevant.
Among these, however, a very large part represents plugins, collections, components, engines, databases, frameworks, libraries, templates, build tools, command line utilities, etc.
Therefore, we hypothesized that the population size for applications would approximately be $1/20$ of this.
Using this as a first rough estimate, we calculated preliminary sample sizes of 375 JS and 341 TS projects.
For each language, we then sorted the lists of retrieved projects by stars.
Starting from the most popular ones, we took batches of 300 projects, which the script then analyzed for selection according to three criteria:

\begin{itemize}
    \item \textbf{Project is an application:} to verify this, the \texttt{README.md} file and project description were automatically examined with the Latent Dirichlet Allocation (LDA) topic modeling algorithm\footnote{\url{https://radimrehurek.com/gensim/models/ldamodel.html}}. LDA takes text as input and outputs topics, where each is a composite of keywords with weights. Topics like \enquote{plugin}, \enquote{module}, \enquote{extension}, \enquote{API}, \enquote{database}, \enquote{framework}, \enquote{library}, etc. were excluded.
    \item \textbf{Project has closed bugs:} to calculate the mean bug resolution time later on, only repositories that contained closed bug issues written in English were included.
    \item \textbf{Project has more than 30 commits:} to ensure reasonable development activity, e.g., avoiding dead projects or projects where all code was pushed within a few commits, we only included repositories with more than 30 commits.
\end{itemize}

All repositories that passed these automated checks were manually examined.
After sorting out non-valid samples, the same process was repeated.
To illustrate the required effort: the script for the TypeScript sampling was running for a total of 14 hours with interruptions.
Until TS selection was finalized, we manually examined over 1,200 repositories, i.e., the \texttt{README.md}, the number of issues, the used languages, as well as the number of commits.
The sampling process also revealed that we had overestimated the population sizes:
out of the 66,000 TypeScript repositories, more than 57,000 were examined by our script.
Of these, 305 met the defined characteristics.
From the more than 300,000 JavaScript repositories, only about 41,000 were examined by the script, with 299 valid samples returned.
Extrapolating this number to 300,000 results in a population size of about 2,200.
Since the available applications became less with the decreasing number of stars, this value could be even smaller.
As a result, our final sample sizes of 299 JS and 305 TS repositories are quite suitable.

Table \ref{tab:StudyObjectValues} presents statistics on the selected projects.
A total of 604 applications with over 16M lines of code, 575,699 commits, and 214,075 issues were examined.
Commits and lines of code were only counted for the corresponding programming languages.
Table \ref{tab:PrimProgLang} shows the distribution of projects over the percentages of the primary programming language.
Among JavaScript applications, 12\% had a backend with a LoC share of more than 10\%.
Similarly, almost 10\% of TypeScript applications had a backend that contained more than 10\% of the project.

\begin{table}[H]
  \centering
%   \small
  \begin{tabular}{lrrrr}
    & Projects & Total LoC & Total Commits & Total Issues \\
    \hline
    \hline
    JS & 299 & 7,496,726 & 235,535 & 56,578 \\
    TS & 305 & 8,683,600 & 340,164 & 157,497 \\
    \hline
    Total & 604 & 16,180,326 & 575,699 & 214,075\\
  \end{tabular}
  \caption{Properties for JavaScript (JS) and TypeScript (TS)}
  \label{tab:StudyObjectValues}
  \vspace{-3mm}
\end{table}

\begin{table}[H]
  \centering
  \small
  \begin{tabular}{lrrrr}
     & PL $ \geq \ $ 90\% & 90\%> PL $ \geq \ $80\% & 80\%> PL $ \geq \ $ 70\% & 70\%> PL $ \geq \ $ 60\% \\
    \hline
    \hline
    JS & 125 & 84 & 49 & 41 \\
    TS & 138 & 82 & 45 & 40 \\
  \end{tabular}
  \caption{Share of the primary programming language (PL)}
  \label{tab:PrimProgLang}
  \vspace{-3mm}
\end{table}

\subsection{Data Collection}
\label{sec:DataCollection}
Similar to the sampling process, collecting the necessary data from the selected projects was also automated using a Python script. 
The two main data sources were the GitHub API and the static analysis tool SonarQube with its API\footnote{\url{https://docs.sonarqube.org/latest/extend/web-api/}}.
Since SonarQube requires the source code, all repositories were downloaded via the GitHub API.
To allow SonarQube the consistent analysis of TS projects, we deleted existing \texttt{tsconfig.json} files, and replaced them with a default file in the root directory.
In addition to the number of code smells and cognitive complexity, we also extracted ncloc from SonarQube, which is the number of lines of code without comments and without empty lines.
Only lines that actually involve JS and TS code were included.

For the bug fix commits, the script also examined if the bug concerns the primary language: only if the file extensions \texttt{.js} (JavaScript), \texttt{.jsx} (React), \texttt{.vue} (Vue.js), \texttt{.ts} (TypeScript), or \texttt{.tsx} (React) were involved, the bug fix commit was included.
Due to the GitHub API rate limiting, this step could not be included in the sampling process, and, as a result, 15 applications with less than 30 commits were not considered for the bug fix commit ratio analysis.

Concerning bug resolution time, the following data was extracted from the GitHub API per issue:
date of opening, date of the last comment, date of closing, title, description, and labels.
During manual inspections, we identified several issues that were opened and directly closed again, potentially to document bugs which had already been fixed.
Therefore, issues that were open for less than two minutes were deleted.
Likewise, issues that were open for more than a year were also removed, as they were probably forgotten and simply closed after a long time.
Projects with less than 5 bug issues were not included in the data collection for this metric.
Afterwards, the mean bug resolution time was calculated per project.

For RQ2, we also needed the number of \texttt{any} types per TypeScript project.
We used ESLint\footnote{\url{https://github.com/typescript-eslint/typescript-eslint}} with the \texttt{typescript-eslint} plugin to collect this metric.
Since using the \texttt{any} type is not recommended, ESLint includes the \texttt{no-explicit-any} rule.
To automate this, the Python script deleted exiting ESLint configurations, and replaced them with a default configuration with only the relevant rule.
Afterwards, ESLint was installed with the TS plugin and executed, and the resulting value was stored.
To avoid project size as a confounder, the number of \texttt{any} types was normalized by dividing by LoC.

\subsection{Statistical Analysis}\label{sec:StatisticalAnalysis}
Since we have a total of eight hypotheses, we needed to counteract the multiple comparisons problem.
With a targeted significance level of $p < 0.05$, we used the Bonferroni correction~\cite{Armstrong2014} to adjust the significance level accordingly, i.e. $0.05 / 8 = 0.00625$.
Thus, if p was less than 0.00625, we rejected the null hypothesis and accepted the alternative.

After examining the data for RQ1 with box plots, we first checked the distribution of our collected metrics to select a fitting hypothesis test.
With a normal distribution, the parametric two-sample t-test could be performed, otherwise, a non-parametric test is needed~\cite{Boslaugh2013}.
We used the Shapiro-Wilk test\footnote{\url{https://docs.scipy.org/doc/scipy/reference/generated/scipy.stats.shapiro.html}} for this, which checks the null hypothesis that the data is not from a normal distribution~\cite{Hanusz2016}.
As a result, none of our metrics were normally distributed.
We therefore selected the non-parametric Mann-Whitney U test\footnote{\url{https://docs.scipy.org/doc/scipy/reference/generated/scipy.stats.mannwhitneyu.html}}, which checks whether two data sets come from the same distribution~\cite{Boslaugh2013}.
For tests where we could reject the null hypothesis, we also calculated Cohen's d as the effect size based on the test statistic (U-Value) following Lenhard and Lenhard~\cite{Lenhard2016}.
According to Sawilowsky~\cite{Sawilowsky2009}, the interpretation of the value is as follows:
\begin{itemize}
    \item \textbf{d < 0.5}: small effect
    \item \textbf{0.5 < d < 0.8}: medium effect
    \item \textbf{d > 0.8}: large effect
\end{itemize}

For RQ2, we started the evaluation with scatter plots.
Afterwards, the statistical relationship and its significance was tested using the Spearman correlation test\footnote{\url{https://docs.scipy.org/doc/scipy/reference/generated/scipy.stats.spearmanr.html}}.
The Pearson correlation test was not applied, as normal distribution is required to assess significance~\cite{Artusi2002}.
Some studies claim that the test is robust to a violation of normality~\cite{Havlicek1976}.
Others, however, are rather critical of this~\cite{Charles1972}.
The test produces a p-value and a correlation coefficient between -1 and 1, where 0 implies no correlation and 1 or -1 an exact monotonic relationship~\cite{Artusi2002}.
A positive number means that if the number of \texttt{any} types per LoC increases, the metric values also increase, while with a negative number, metric values would decrease.

\section{Results}
Following the outlined methodology, we present the concrete results per RQ and hypothesis.
At the end of the section, we summarize all test statistics in Table~\ref{tab:results} for a better overview.
Aggregated values shown in tables, i.e., mean and median, are calculated using the size-normalized values per project, and can therefore differ from the respective aggregate using the shown totals per language.

\subsection{Software Quality Differences (RQ1)}
For RQ1, we directly compared the outlined software quality characteristics between our JavaScript and our TypeScript samples.
With \textbf{H$_{1}^{1}$}, we investigated whether TypeScript applications show significantly less code smells per LoC.
Table~\ref{tab:CodeSmells} presents the aggregated metrics for this question.
In total, 313,098 code smells were detected in the 604 applications.
JavaScript has more than twice as many code smells compared to TypeScript, while simultaneously having over one million \textit{fewer} lines of code.
As a result, JS apps have, on average, roughly 12 code smells per 1 kLoC more than TS apps.

\begin{table}[H]
    \centering
    \begin{tabular}{lrrrrr}
        & Projects & LoC & Code Smells & Mean & Median\\
       \hline
       \hline
       JS & 299 & 7,496,726 & 217,957 & 0.025545 & 0.020132\\
       TS & 305 & 8,683,600 & 95,132 & 0.013368 & 0.011143\\
    \end{tabular}
    \caption{Code smells for JavaScript (JS) and TypeScript (TS)}
    \label{tab:CodeSmells}
    \vspace{-3mm}
\end{table}

When verifying the significance of this difference with the Mann-Whitney U test, we receive a p-value of 2.7e-15, i.e., much smaller than the required significance level of 0.00625.
Therefore, the null hypothesis can be rejected, and we accept the alternative.
Cohen's d yields a value of 0.671, which indicates a medium effect size.

\begin{tcolorbox}
\textbf{Result for H$_{1}^{1}$:} TypeScript applications have significantly less code smells than JavaScript applications, i.e., better code quality.
\end{tcolorbox}

For \textbf{H$_{1}^{2}$}, we similarly tested if TypeScript applications have significantly lower cognitive complexity (CC) per LoC.
An overview of the data collected is given in Table~\ref{tab:CogCompl}.
With over 16M lines of code, the total cognitive complexity score was over 4M.
This time, JS applications have nearly five times the summarized cognitive complexity of TS apps, with 1M lines of code less.
TypeScript has a mean CC of around 90 per kLoC, whereas JavaScript has 322 per kLoC, has more than three times as much.

\begin{table}[H]
    \centering
    \begin{tabular}{lrrrrr}
         & Projects & LoC & CC & Mean & Median\\
        \hline
        \hline
        JS & 299 & 7,496,726 & 3,435,760 & 0.321999 & 0.157013\\
        TS & 305 & 8,683,600 & 722,687 & 0.089552 & 0.077416
    \end{tabular}
    \caption{Cognitive complexity (CC) per LoC}
    \label{tab:CogCompl}
    \vspace{-3mm}
\end{table}

With such pronounced mean and median differences, the Mann-Whitney U test resulted in a p-value of 5.14e-18, again substantially lower than the significance level of 0.00625.
We therefore rejected the null hypothesis and accepted the alternative.
However, Cohen's d only yields a value of 0.339, which indicates a small effect size.

\begin{tcolorbox}
\textbf{Result for H$_{1}^{2}$:} TypeScript applications have significantly lower cognitive complexity than JavaScript applications, i.e., better code understandability.
\end{tcolorbox}

For \textbf{H$_{1}^{3}$}, we tested if TypeScript applications have a significantly lower bug fix commit ratio.
The data for this analysis is summarized in Table~\ref{tab:BugFixCom}.
As described in Section~\ref{sec:DataCollection}, not all projects could be used due to excluding several commits after the sampling.
A total of 575,699 JS and TS commits, of which 97,205 contained a bug-fix, were examined.
TypeScript apps have over 40\% more commits in total, but also more than twice as many bug fixes.
JavaScript applications therefore average roughly one bug fix for every 8 commits, while TypeScript applications do so for every 5 commits, i.e., over 60\% more.

\begin{table}[H]
    \centering
    \begin{tabular}{lrrrrr}
         & Projects & Commits & BF Commits & Mean & Median\\
    \hline
    \hline
    JS & 285 & 235,535 & 30,717 & 0.126364 & 0.109375\\
    TS & 288 & 340,164 & 66,488 & 0.206436 & 0.163461\\
    \end{tabular}
  \caption{Bug fix (BF) commits for JavaScript and TypeScript}
  \label{tab:BugFixCom}
  \vspace{-3mm}
\end{table}

The lower aggregates for JavaScript already indicate that the hypothesis will not hold.
As expected, the Mann-Whitney U test results in a p-value of 0.99.
Therefore, the null hypothesis cannot be rejected.

\begin{tcolorbox}
\textbf{Result for H$_{1}^{3}$:} TypeScript applications have a higher bug fix commit ratio than JavaScript applications, i.e., are more or equally prone to bugs.
\end{tcolorbox}

Lastly, \textbf{H$_{1}^{4}$} analyzes if it takes significantly less time in TypeScript applications to close bug issues, i.e., if these projects require less bug resolution time.
The aggregated data is shown in Table~\ref{tab:BugResul}.
As with the bug fix commit ratio, we could not use the complete data set for the analysis.
Of the 214,075 collected issues, 67,711 contain a bug, which is substantially less than the number of bug fix commits, especially for JavaScript.
However, TypeScript projects not only have more documented issues in general, but also more than three times as many bug issues as JavaScript projects.
Concerning bug resolution time, JavaScript developers needed on average 31.86 days to close a bug issue, which is over a day less than TypeScript developers.
This difference is even more pronounced with the median.

\begin{table}[H]
    \centering
    \begin{tabular}{lrrrr}
         & Projects & Bug Issues & Mean BR Time & Median BR Time\\
        \hline
        \hline
        JS & 183 & 15,894 & 31.86 days  & 25.59 days\\
        TS & 245 & 51,817 & 33.04 days & 28.49 days\\
    \end{tabular}
    \caption{Bug resolution (BR) time per language}
    \label{tab:BugResul}
    \vspace{-3mm}
\end{table}

The differences in aggregation measures again indicate that the hypothesis will not hold.
Understandably, the Mann-Whitney U test provides a p-value of 0.75.
Therefore, the null hypothesis cannot be rejected.

\begin{tcolorbox}
\textbf{Result for H$_{1}^{4}$:} TypeScript applications take more or equal time in bug resolution than JavaScript applications.
\end{tcolorbox}

\subsection{Effect of the \texttt{any} Type in TypeScript (RQ2)}
With RQ2, we wanted to test if there are significant correlations between each of the four software quality characteristics and the frequency of \texttt{any} types in TypeScript applications.
To get a general overview of the data, a scatter plot was created for each of the four hypotheses.
However, since the Spearman test converts the values into ranks, the correlation coefficient is difficult to interpret geometrically.
Table \ref{tab:anyTypes} shows the collected data on \texttt{any} type usage.
For the 305 TypeScript projects, ESLint reported a total of 79,735 \texttt{any} type rule violations, i.e., on average roughly 261 per project.
However, this value does not control for project size.
Normalizing with LoC, the average TS app uses the \texttt{any} type once for every 100 lines of code.

\begin{table}[H]
    \centering
    \begin{tabular}{lrrrrr}
         & Min & Max & Mean & Median & Total\\
        \hline
        \hline
        \texttt{any} types & 0 & 5326 & 261.43 & 70 & 79,735\\
        \texttt{any} types per LoC & 0 & 0.1063 & 0.0103 & 0.0072 & --
    \end{tabular}
    \caption{Descriptive statistics on \texttt{any} type usage}
    \label{tab:anyTypes}
    \vspace{-3mm}
\end{table}

\begin{figure}
    \center
    \includegraphics[width=\columnwidth]{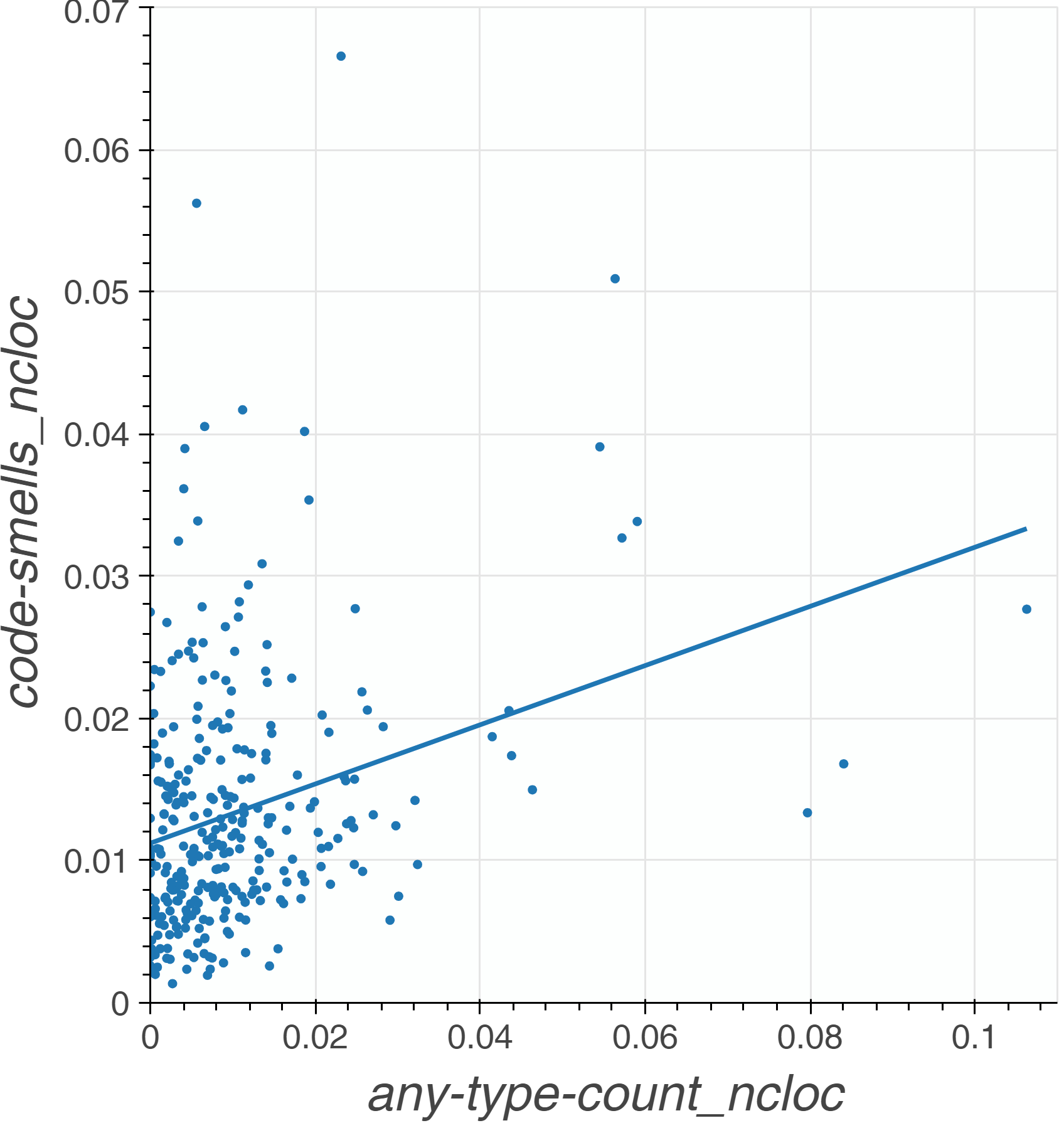}
    \caption{Code smells vs. \texttt{any} types (H$_{1}^{5}$)}
    \label{fig:ScatterCodeSmells2}
\end{figure}

For \textbf{H$_{1}^{5}$}, we tested if there is a significant positive correlation between the number of \texttt{any} types per LoC and the number of code smells per LoC in TypeScript applications, i.e., if more \texttt{any} types go together with more code smells.
An overview of the data for this is presented as a scatter plot in Fig.~\ref{fig:ScatterCodeSmells2}.

The figure seems to suggest a weak correlation, which is confirmed by the Spearman correlation test, resulting in a p-value of 2.5e-06 and a correlation coefficient of 0.26.
While the p-value is lower than the specified significance level of 0.00625, the coefficient indicates only a weak strength for the relationship.
Nonetheless, we rejected the null hypothesis and accepted the alternative.

\begin{tcolorbox}
\textbf{Result for H$_{1}^{5}$:} The frequency of using the \texttt{any} type shows a positive but weak correlation with the number of code smells in TypeScript applications ($\rho = 0.26$).
\end{tcolorbox}

\begin{figure}
    \center
    \includegraphics[width=\columnwidth]{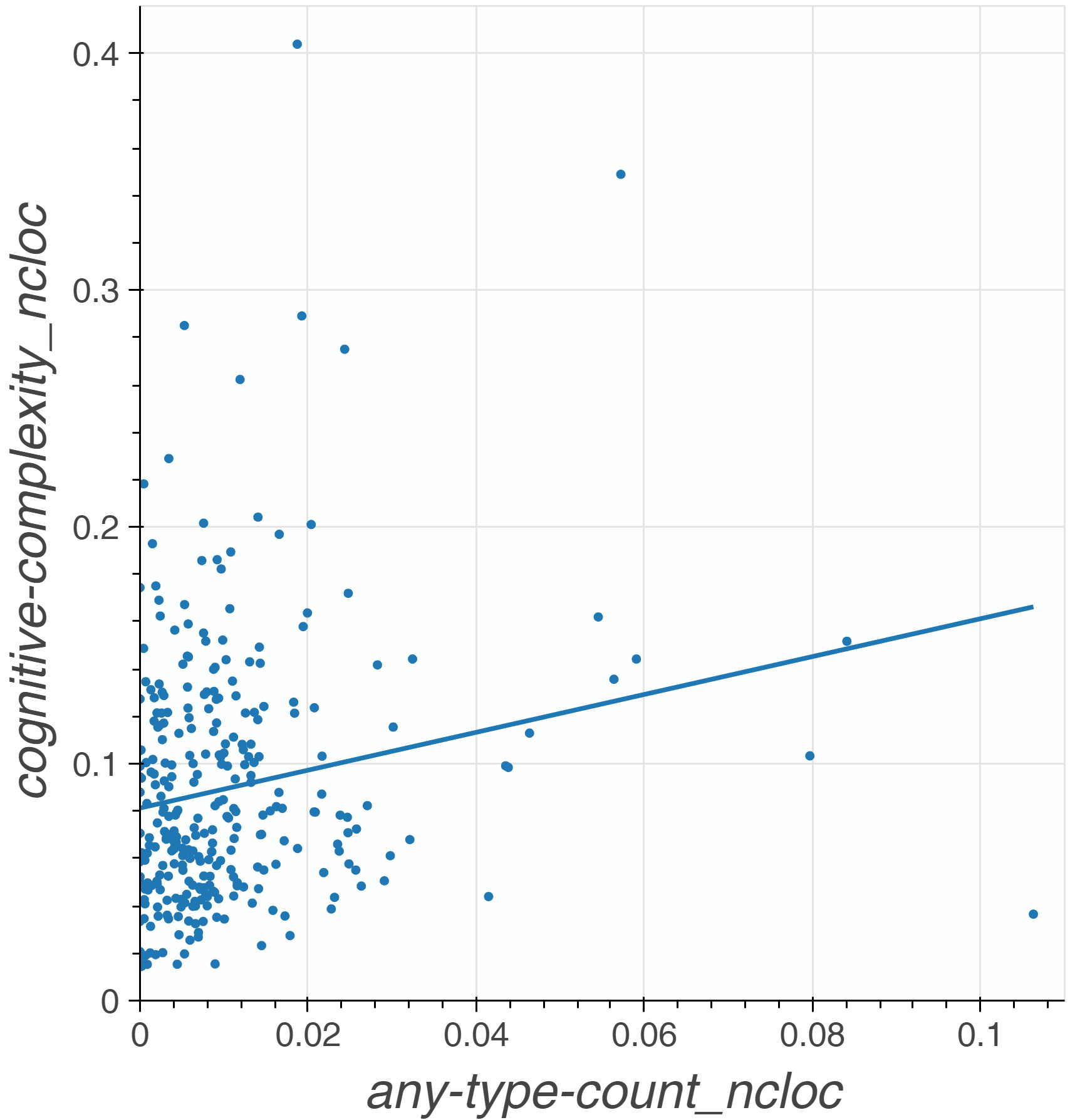}
    \caption{Cognitive complexity vs. \texttt{any} types (H$_{1}^{6}$)}
    \label{fig:ScatterCodeUnder2}
\end{figure}

For \textbf{H$_{1}^{6}$}, we tested if there is a significant positive correlation between the number of \texttt{any} types per LoC and cognitive complexity per LoC.
The collected data is visualized as a scatter plot in Fig.~\ref{fig:ScatterCodeUnder2}.
Again, the scatter plot seems to suggest a small relationship.
The Spearman correlation test resulted in a p-value of 4.7e-04 and a correlation coefficient of 0.19.
The p-value is significant at the 0.00625 level, even though the coefficient indicates only a weak correlation between the two metrics.
Still, we rejected the null hypothesis and accepted the alternative.

\begin{tcolorbox}
\textbf{Result for H$_{1}^{6}$:} The frequency of using the \texttt{any} type shows a positive but weak correlation with cognitive complexity in TypeScript applications ($\rho = 0.19$).
\end{tcolorbox}

\begin{figure}
    \center
    \includegraphics[width=\columnwidth]{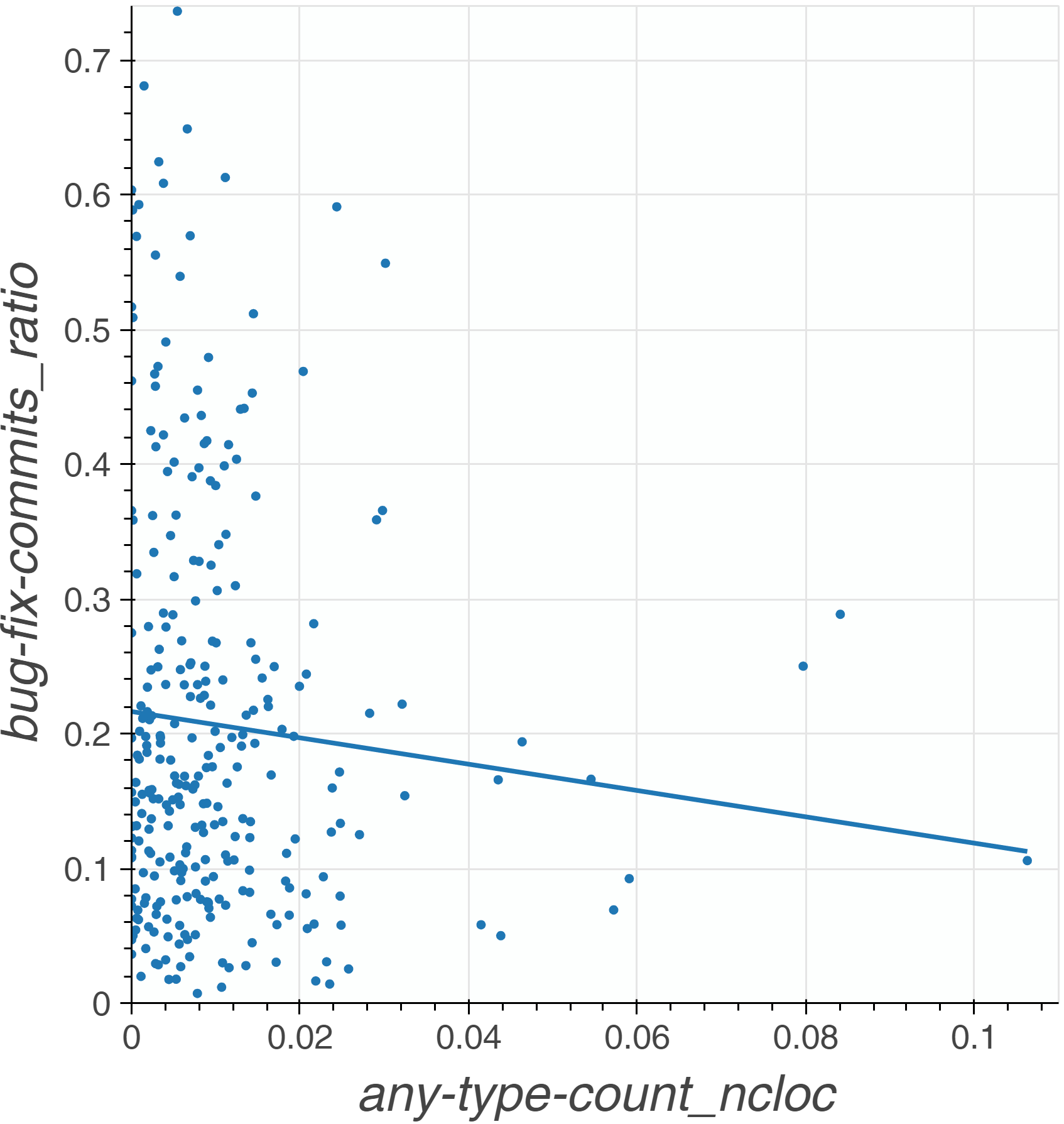}
    \caption{Bug fix commit ratio vs. \texttt{any} types (H$_{1}^{7}$)}
    \label{fig:ScatterBugFix2}
\end{figure}

For \textbf{H$_{1}^{7}$}, we tested if there is a significant positive correlation between the number of \texttt{any} types per LoC and the bug fix commit ratio, i.e., if ignoring type safety increases bug proneness.
Again, we created a scatter plot for a quick overview of the data, which can be seen in Fig.~\ref{fig:ScatterBugFix2}.
This time, however, the scatter plot suggest a slight negative relationship, i.e., the \textit{opposite} of what we expected.
Performing the Spearman correlation test confirms this, with a p-value of 0.77 and a coefficient of -0.04.
Therefore, there is not enough evidence to reject the null hypothesis.

\begin{tcolorbox}
\textbf{Result for H$_{1}^{7}$:} There is no significant correlation between the frequency of using the \texttt{any} type and the bug fix commit ratio in TypeScript applications.
\end{tcolorbox}

\begin{figure}
    \center
    \includegraphics[width=\columnwidth]{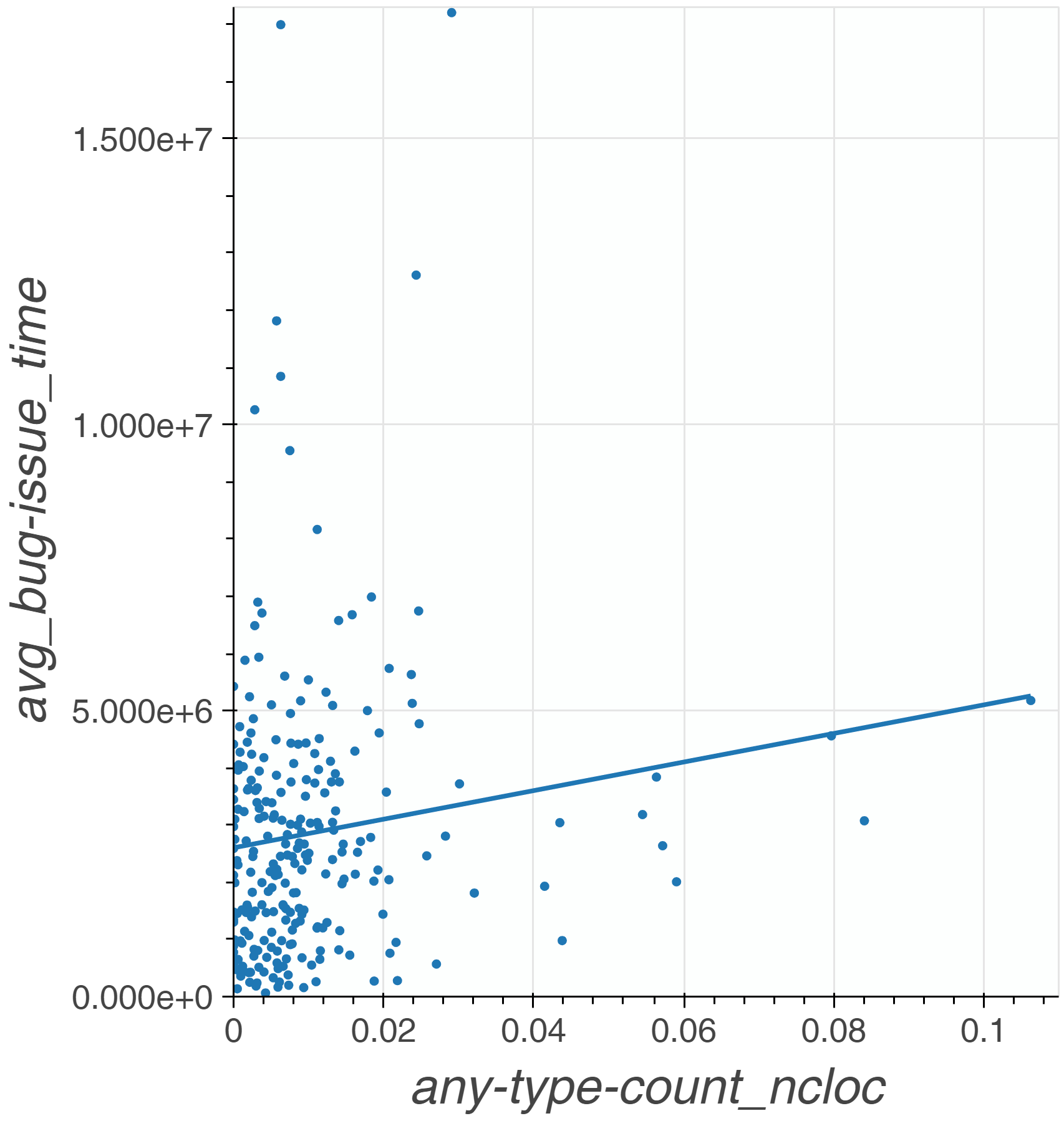}
    \caption{Bug resolution time vs. \texttt{any} types (H$_{1}^{8}$)}
    \label{fig:ScatterBugResol2}
\end{figure}

Lastly, \textbf{H$_{1}^{8}$} tested if there is a significant positive correlation between the number of \texttt{any} types per LoC and the bug resolution time, i.e., if bugs take longer to fix if type safety is ignored.
The scatter plot to illustrate the collected data is presented in Fig.~\ref{fig:ScatterBugResol2}.

In this case, the scatter plot hints at a small positive relationship.
The Spearman correlation test resulted in a p-value of 0.0034 and a correlation coefficient of 0.17.
This is still significant at the 0.00625 level.
Once again, however, the coefficient indicates only a weak strength for the relationship.
The null hypothesis was rejected and the alternative accepted.

\begin{tcolorbox}
\textbf{Result for H$_{1}^{8}$:} The frequency of using the \texttt{any} type shows a positive but weak correlation with bug resolution time in TypeScript applications ($\rho = 0.17$).
\end{tcolorbox}

\subsection{Summary}
In summary, we could accept two of the four hypotheses for RQ1, namely the ones for code quality (H$_{1}^{1}$) and code understandability (H$_{1}^{2}$), with a medium and small effect size respectively.
However, both bug-related hypotheses (H$_{1}^{3}$ and H$_{1}^{4}$) did not lead to significant results.
For RQ2, three of the four hypotheses could be accepted, with the one for bug proneness (H$_{1}^{7}$) being the exception.
The strength for all significant correlations was weak, though.
Table~\ref{tab:results} summarizes these results.

\begin{table*}[ht]
    \centering
    \begin{tabular}{lp{9cm}rrrr}
        RQ & Alternative Hypothesis & Test Statistic (U) & p-Value  & Cohen's d or $\rho$ & Accepted\\
        \hline
        \hline
        \multirow{5}{*}{RQ1} &
        \textbf{H$_{1}^{1}$}: TS applications exhibit better code quality than JS applications. & 28842.5 & \textbf{2.78e-15} & 0.671 & Yes\\
        & \textbf{H$_{1}^{2}$}: TS applications exhibit better code understandability than JS applications. & 27219.0 & \textbf{5.14e-18} & 0.339 & Yes\\
        & \textbf{H$_{1}^{3}$}: TS applications are less prone to bugs than JS applications. & 54395.5 & 0.99 & -- & No\\
        & \textbf{H$_{1}^{4}$}: TS applications require less bug resolution time than JS applications. & 23281.0 & 0.75 & -- & No\\
        \hline
        \multirow{8}{*}{RQ2} &
        \textbf{H$_{1}^{5}$}: The \texttt{any} type frequency correlates negatively with code quality in TS applications. & -- & \textbf{2.48e-06} & 0.26 &  Yes\\
        & \textbf{H$_{1}^{6}$}: The \texttt{any} type frequency correlates negatively with code understandability in TS applications. & -- & \textbf{4.7e-04} & 0.19 & Yes\\
        & \textbf{H$_{1}^{7}$}: The \texttt{any} type frequency correlates positively with bug proneness in TS applications. & -- & 0.76 & -0.04 & No\\
        & \textbf{H$_{1}^{8}$}: The \texttt{any} type frequency correlates positively with bug resolution time in TS applications. & -- & \textbf{0.0034} & 0.17 & Yes
    \end{tabular}
    \caption{Summary of hypothesis testing results (significant p-values in bold, $\alpha = 0.00625$)}
    \label{tab:results}
\end{table*}

\section{Discussion}
In this section, we offer potential explanations for our results, and discuss their implications.
In both RQs, we could show that TypeScript applications and refraining from using the \texttt{any} type are connected to improved code quality and understandability.
The question remains, however, if our data is strong enough to suggest a causal relationship.
One argument in this direction could be that \texttt{tsc} warns about several facets of erroneous and unoptimized code, and thereby combats the dynamic and flexible nature of JavaScript.
Theoretically, this could have an effect on reducing some code smells, but still is unlikely to account for many others or for reducing cognitive complexity.

Another explanation could be that TypeScript may be more frequently chosen by experienced developers, who may be less likely to introduce code smells and complex code.
Experienced developers may also use static analysis tools to identify and remove code smells more frequently.
Many of the TypeScript projects in our sample used ESLint, so the assumption that several also use SonarQube or other static analyzers may be reasonable.
Another indicator for an experience difference could be the much greater popularity of TS projects in our sample:
TS apps averaged roughly 2.5k stars (759,619 in total), while JS apps averaged only 604 (180,690 in total).
Despite TS outliers like Storybook\footnote{\url{https://github.com/storybookjs/storybook}} (68k stars), code-server\footnote{\url{https://github.com/coder/code-server}} (51k), or Grafana\footnote{\url{https://github.com/grafana/grafana}} (47k), the median difference is even more pronounced (84 vs. 738 stars).
In theory, increased popularity could have the effect of attracting more skilled or experienced contributors.
Nonetheless, JavaScript apps in our sample were from the top 15\% of popular JS projects and also supported by numerous contributors, making it difficult to believe that their developers may be significantly less skilled than their TS counterparts.
Unfortunately, it is not really possible to directly control for the maturity or experience of the development teams with our study design.

Contrary to expectations, only a single one of our bug-related hypotheses could be accepted, namely that a decreased usage of the \texttt{any} type is correlated with a faster bug resolution time in TypeScript apps.
In comparison to JavaScript apps, TS projects in our sample were \textit{more} bug prone and required \textit{more} time to fix bugs.
Moreover, a decreased usage of the \texttt{any} type was \textit{not} correlated with fewer bugs.
These results are counterintuitive to the theory that TS developers are more skilled and experienced, as more experienced developers should produce \textit{fewer} bugs.
Project popularity, i.e., stars, could be a confounder for bug proneness (more users will find and report more bugs), but we control for this by using the bug fix commit ratio, not simply the number of bugs.
In our sample, there is no correlation between the bug fix commit ratio and number of stars (Pearson's r of 0.03, p-value of 0.42), so popularity also fails as an explanation.
The results also directly contradict Gao et al.~\cite{Gao2017}, who found that TypeScript can prevent 15\% of JavaScript bugs, and Zhang et al.~\cite{Zhang2020}, who found that dynamically typed languages require 60\% more time for bug resolution than statically typed languages.
However, both studies did not solely focus on applications, and the latter also did not include TypeScript in their data set.

% Could TS projects simply be inherently more complex?
One potential explanation could be that TypeScript is chosen more often for projects with very high inherent domain complexity, which could counteract any potential benefits for preventing and fixing bugs.
Domain complexity is very hard to measure, though, and we know that TS projects at least have significantly lower accidental code complexity.
If we use size as a suboptimal approximation of domain complexity, we see that the median LoC for our JS sample is 7.2k, while it is 11.2k for TS, i.e., more than 55\% larger.
While TypeScript with its type definitions is definitely more verbose, it seems unlikely that this alone would account for such a large difference.
However, even if we assume the increased domain complexity hypothesis to be true, this still would not explain why reduced usage of the \texttt{any} type is not correlated with fewer bugs in our TS sample.

% Could TS developers simply be more thorough in documenting bugs?
A second potential explanation could be that TypeScript developers simply were more thorough in documenting bugs and their fixes than their JS counterparts.
Both the number of commits and the number of issues are substantially higher in our TS sample, thereby presumably supporting this notion.
That JavaScript developers committed less could also indicate that they sometimes fixed multiple bugs in one commit, even though this is unlikely to result in such large discrepancies.
As before, however, this theory would also not sufficiently explain the missing correlation between the degree of type safety and bug proneness in our TS sample.

In summary, the implications of our results are that using TypeScript and adhering to its type safety is associated with increased code quality and understandability, even though it is unlikely that this is directly and entirely caused by TypeScript.
Based on the comparison, JS developers should be more conscious about code quality.
However, the presumed benefits of TypeScript for preventing and quickly resolving bugs could not be verified in this study, which hints at a more complicated interaction mechanism.
At least in our sample of web applications, TS seems not as superior for preventing bugs as often believed, and JS does not need to hide here.
If anything, TS developers should not fall into a false sense of security.
Lastly, while refraining from using the \texttt{any} type has some benefits, it does not seem to be significantly related to bug prevention, i.e., TS developers do not have to try to religiously avoid it at all cost.
The effort may not be worth it in several cases.
More research on this is needed, though.

\section{Threats to Validity}
Even though we automated major parts of this study, conducted extensive manual quality checks, and adopted best practices from existing similar studies, several threats to validity remain.
One area for potential threats is an \textbf{insufficient or non-representative sample}.
Since the real population size is unknown, we roughly estimated it during the sampling procedure, and later adjusted our estimates as we learned more about our sample.
With roughly 300 projects per language, our sample size is comparable to or larger than the majority of related work.
To strengthen the internal validity by making the study objects more comparable (web apps are fairly homogeneous), we consciously excluded frameworks, libraries, or build tools.
This also reduced the population size considerably, but simultaneously sacrificed  a certain degree of external validity.
Moreover, it is still possible that we falsely excluded several relevant projects during the automated parts of the sampling, e.g., during the LDA analysis of \texttt{README.md} and project description, thereby unnecessarily reducing our sample size.

Projects were selected based on popularity, i.e., GitHub stars, which is a frequently used, albeit imperfect way of sampling.
For TypeScript, the popularity of projects was quite diverse.
For Java\-Script, however, we only examined the top 15\% of repositories by popularity.
This is not representative of all applications.
Nevertheless, we believe this to be a better approach than to randomly select projects.
The lower the popularity of the projects, the fewer applications also matched our requirements for inclusion.
Random sampling would have therefore extended the required time and effort quite substantially.

What cannot be fully avoided is the proportion of programming languages that are not JavaScript or TypeScript.
As we specifically focused on applications, basically all of them contained HTML, CSS, and sometimes also a backend language.
For JavaScript and TypeScript, about 10\% of projects also contained a backend that accounted for more than 10\% of the total LoC.
This amount should be small enough to not substantially affect the data set, plus backend languages were filtered out during data collection, e.g., for commits.
Excluding all applications with a backend would have restricted the population size even further.

Our entire sample consists of open-source frontend applications.
Great care must therefore be taken when generalizing our findings to other types of systems, e.g., commercial applications.
Moreover, the number of developers and their experience was not taken into account, as it would have been difficult if not impossible to control for.
The size distributions of the projects were also not considered during sampling, even though we controlled for size as a confounding variable.

The second prominent area for threats is \textbf{data collection}.
Our operationalized facets of software quality (code smells, cognitive complexity, bug proneness, and bug resolution time) are all frequently used, with empirical support for their effectiveness.
Nonetheless, there are obviously other facets of software quality that we did not cover in this study.

As proposed by Zhang et al.~\cite{Zhang2020}, we automated the detection of bug fix commits by searching for \enquote{bug} and \enquote{fix} in commit messages.
As quality control, we manually examined 0.5\% (500 randomly sampled commits) of over 97,000 bug fix commits in our sample to determine the false positive rate and how often more than one bug fix was included.
6\% of commits contained no bug fix, and 8\% contained more than one.
However, these numbers are rather low and should not have a major impact on the results, especially when assuming that this affects both JS and TS to a certain degree.

Similar inaccuracies may apply to the measuring of bug resolution time.
Instead of using the closing timestamp of the bug issue, we opted for the timestamp of the last comment.
Zheng et al.~\cite{Zheng2015} reported this as the more accurate method, and it has since been adopted in other studies, e.g., by Zhang et al.~\cite{Zhang2020}.
Nonetheless, it is still reasonable to assume a small margin for error within our bug resolution time measurements, even though we do not think that results are impacted by this.

Moreover, some bug issues for the backend parts may have been included, as there is no way to automatically identify them reliably.
The approximately 10\% of applications that have a backend were considered low enough that they would not affect the dataset significantly.

For the detection of the number of used \texttt{any} types, ESLint with the \texttt{typescript-eslint} plugin used, whereas SonarQube was used for code smells, cognitive complexity, and LoC.
To exclude errors, projects with strange values were manually double-checked, and, in rare cases, the analysis was re-run if necessary.
Nonetheless, it is possible that smaller inaccuracies could have occurred and remained undetected in the data set, as we could not check every measurement.
However, such errors should be sufficiently rare to not significantly affect the results.

Lastly, some threats for the \textbf{conclusion validity} of this study remain.
Correlation does not automatically mean causation, and even though we eliminated several potential confounders, our results still have to be interpreted with great care.
As an example, it does not follow that using JavaScript \textit{causes} fewer bugs than using TypeScript.
We simply showed that, in a direct comparison between JS and TS apps and contrary to popular belief, TypeScript was not significantly related to fewer bugs.
However, more research is needed to support and explain (or refute) our findings, for which we hopefully created a useful foundation.

\section{Conclusion}
Since TypeScript (TS) is assumed to provide benefits for software quality in comparison to JavaScript (JS), yet there is insufficient empirical evidence for this claim, we conducted a repository mining study to analyze and compare a large number of open-source applications on GitHub.
We examined a total of 604 repositories (299 JS, 305 TS) with over 16M lines of code. 
Statistical analysis revealed that TypeScript applications show significantly better code quality and understandability than JavaScript applications.
Surprisingly, however, bug proneness and bug resolution time were both lower for JavaScript projects, thereby defying the assumed benefits of static typing in this area.
Within our TypeScript sample, insisting on type safety by refraining from extensively using the \texttt{any} type showed significant, albeit small correlations with better software quality, except for, again, bug proneness.
Potential explanations include an increased domain complexity of our TypeScript sample or a tendency of TS developers to document bugs more thoroughly, even though both are not fully convincing.
Our findings therefore indicate that using TypeScript and adhering to its type safety has benefits, but not the straightforward advantage of directly reducing bugs.
More research in this area is needed, though.
Future studies should try to replicate or dispute our findings, and especially try to shed light on the potential influence of system complexity and developer experience.
Additionally, conducting a similar study with a broader focus than client-side applications will enable interesting comparisons.
For transparency and to enable such replications, we share our study artifacts on Zenodo\footnote{\url{https://doi.org/10.5281/zenodo.5886595}}.

% \begin{acks}
% \end{acks}

\bibliographystyle{ACM-Reference-Format}
\bibliography{references}

\end{document}